\documentclass{aa}
\usepackage{graphicx,natbib,times,bm,url}
\usepackage{amssymb}
\graphicspath{{./fig/}{./png/}}
\newcommand{\EQ}{\begin{equation}}
\newcommand{\EN}{\end{equation}}
\newcommand{\EQA}{\begin{eqnarray}}
\newcommand{\ENA}{\end{eqnarray}}
\newcommand{\eq}[1]{(\ref{#1})}

\newcommand{\Eq}[1]{Eq.~(\ref{#1})}

\newcommand{\Fig}[1]{Fig.~\ref{#1}}

\newcommand{\Figs}[2]{Figs.~\ref{#1} and \ref{#2}}

\newcommand{\Tab}[1]{Table~\ref{#1}}

{}
{}
{}

{}
{}
{}
{}
{}
{}
{}
{}
{}
\newcommand{\meanBB}{\overline{\mbox{\boldmath $B$}}{}}{}
{}
{}
{}
{}
{}
{}
{}
{}
{}
{}
{}

{}
{}
{}

\newcommand{\meanB}{\overline{B}}

\newcommand{\meanU}{\overline{U}}

{}

{}
{}

\newcommand{\tvA}{\tilde{v}_{\rm A}}

\newcommand{\eee}{\hat{\mbox{\boldmath $e$}} {}}

\newcommand{\xxx}{\hat{\mbox{\boldmath $x$}} {}}
\newcommand{\yyy}{\hat{\mbox{\boldmath $y$}} {}}

\newcommand{\kk}{\bm{k}}

\newcommand{\xx}{\bm{x}}

\newcommand{\BB}{\bm{B}}

\newcommand{\UU}{\mbox{\boldmath $U$} {}}

\newcommand{\JJ}{\mbox{\boldmath $J$} {}}

\newcommand{\AAA}{\mbox{\boldmath $A$} {}}

\newcommand{\ff}{\mbox{\boldmath $f$} {}}

\newcommand{\nab}{\mbox{\boldmath $\nabla$} {}}
\newcommand{\OO}{\bm{\Omega}}

\newcommand{\SSSS}{\mbox{\boldmath ${\sf S}$} {}}

\newcommand{\DD}{{\rm D} {}}
\newcommand{\DDD}{{\cal D} {}}

\def\Co{\mbox{\rm Co}}

\def\Rm{\mbox{\rm Re}_M}

\def\Rey{\mbox{\rm Re}}

\def\Co{\mbox{\rm Co}}
\def\Lu{\mbox{\rm Lu}}

\def\cs{c_{\rm s}}

\def\etafix{\eta_{\rm fixed}}

\def\vA{v_{\rm A}}
\def\VA{V_{\rm A}}

\def\kf{k_{\rm f}}
\def\kmax{k_{\rm max}}

\def\Brms{B_{\rm rms}}

\def\urms{u_{\rm rms}}

\def\brms{b_{\rm rms}}

\def\nut{\nu_{\rm t}}

\def\etat{\eta_{\rm t}}
\def\etatMRI{\eta^{\rm MRI}_{\rm t}}

\def\etatz{\eta_{\rm t0}}

\def\etaT{\eta_{\rm T}}

\def\half{{\textstyle{1\over2}}}

\def\onethird{{\textstyle{1\over3}}}

\newcommand{\yaj}[3]{ #1, {AJ,} {#2}, #3}

\newcommand{\ymonber}[3]{ #1, {Monats.\ Dt.\ Akad.\ Wiss.,} {#2}, #3}
\newcommand{\yapj}[3]{ #1, {ApJ,} {#2}, #3}

\newcommand{\yapjl}[3]{ #1, {ApJ,} {#2}, #3}
\newcommand{\yapjs}[3]{ #1, {ApJS,} {#2}, #3}

\newcommand{\yan}[3]{ #1, {Astron.\ Nachr.,} {#2}, #3}

\newcommand{\yana}[3]{ #1, {A\&A,} {#2}, #3}

\newcommand{\ygafd}[3]{ #1, {Geophys.\ Astrophys.\ Fluid Dyn.,} {#2}, #3}

\newcommand{\ypf}[3]{ #1, {Phys.\ Fluids,} {#2}, #3}

\newcommand{\ysovl}[3]{ #1, {Sov.\ Astron.\ Lett.,} {#2}, #3}

\newcommand{\yprl}[3]{ #1, {Phys.\ Rev.\ Lett.,} {#2}, #3}

\newcommand{\ymn}[3]{ #1, {MNRAS,} {#2}, #3}

\newcommand{\ypre}[3]{ #1, {Phys.\ Rev.\ E,} {#2}, #3}

\newcommand{\yjour}[4]{ #1, {#2}, {#3}, #4}

\newcommand{\ybook}[3]{ #1, {#2} (#3)}

\titlerunning{Effect of turbulent magnetic diffusion on the MRI}
\title{Quantifying the effect of turbulent magnetic diffusion on
the growth rate of the magneto-rotational instability}

\authorrunning{M. S.\ V\"ais\"al\"a et al.}
\author{M. S.\ V\"ais\"al\"a\inst{1,2}, A. Brandenburg\inst{2,3},
Dhrubaditya Mitra\inst{2},
P. J.\ K\"apyl\"a\inst{1,2} \and M. J.\ Mantere\inst{1,4}
}

\institute{
Department of Physics, Gustaf H\"allstr\"omin katu 2a, PO Box 64,
FI-00014 University of Helsinki, Finland
\and
Nordita, KTH Royal Institute of Technology and Stockholm University,
Roslagstullsbacken 23, 10691 Stockholm, Sweden
\and
Department of Astronomy, AlbaNova University Center,
Stockholm University, 10691 Stockholm, Sweden
\and
Aalto University, Department of Information and Computer Science,
PO Box 15400, FI-00076 Aalto, Finland
}
\date{\today,~ $ $Revision: 1.157 $ $}
\begin{document}

\abstract{
In astrophysics, turbulent diffusion is often used in place of
microphysical diffusion to avoid resolving the small scales.
However, we expect this approach to break down when time and length
scales of the turbulence become comparable with other relevant
time and length scales in the system.
Turbulent diffusion has previously been applied to the magneto-rotational
instability (MRI), but no quantitative comparison of growth rates at
different turbulent intensities has been performed.
}{
We investigate to what extent turbulent diffusion can be used
to model the effects of small-scale turbulence on the kinematic
growth rates of the MRI, and how this depends on angular velocity
and magnetic field strength.
}{
We use direct numerical simulations in three-dimensional shearing
boxes with periodic boundary conditions in the spanwise direction
and additional random plane-wave volume forcing to drive
a turbulent flow at a given length scale.
We estimate the turbulent diffusivity using a mixing length
formula and compare with results obtained with the
test-field method.
}{
It turns out that the concept of turbulent diffusion is remarkably
accurate in describing the effect of turbulence on the growth rate of
the MRI.
No noticeable breakdown of turbulent diffusion has been found, even
when time and length scales of the turbulence become comparable
with those imposed by the MRI itself.
On the other hand, quenching of turbulent magnetic diffusivity
by the magnetic field is found to be absent.
}{
Turbulence reduces the growth rate of the MRI in a way that is
the same as microphysical magnetic diffusion.
}
\keywords{turbulence --  magnetohydrodynamics (MHD) -- hydrodynamics }

\maketitle

\section{Introduction}
A cornerstone in the study of astrophysical fluids is linear stability theory \citep{Cha61}.
An important example is the magneto-rotational instability \citep[MRI, see][]{BH98}, 
which will also be the focus of the present paper.
However, the issue is more general, and there are other
instabilities that we mention below.
When studying linear stability, one typically considers a stationary
solution  of the full nonlinear equations, linearizes the equations
about this solution, and looks for the temporal behavior of small
perturbations (wavenumber $k$)  proportional to $e^{\lambda t}$, where $t$ is time
and $\lambda (k)$ is generally complex.
The real part of $\lambda$ is the growth rate, and $\lambda$ as a function
of $k$ is the dispersion relation. 
Linear stability theory is useful to explain
why many astrophysical flows are turbulent
(e.g., accretion disks through the MRI or the stellar convection zones
through the convective instability).

Linear stability theory is also generalized to study the
formation of large-scale instabilities in the presence of turbulent flows; e.g.,
studies of stability of the solar tachocline where convective turbulence
is expected to be present \citep{ASR07,MGD07}.
Let us first revisit this generalization. 
In the case of a turbulent flow, there is no stationary state in the usual
sense; we can at best expect a \textit{statistically} steady state.
In such a situation, the prescription is to average over, or coarse-grain, the
fundamental nonlinear equations (e.g., equations of magnetohydrodynamics) to 
write a set of effective equations valid for large
length and time scales.  
Typical examples of such averaging include Reynolds averaging \citep{Mof78,KR80},
the multiscale techniques \citep{Zhe12}, and application of the dynamical renormalization
group, see, e.g.,~\cite{Gol92}. 
The effective equations themselves depend on the averaging process, and also 
on the length and time scales to which they are applied. 
The averaging process can give rise to new terms in the effective equations 
and it introduces new transport coefficients that are often called turbulent
transport coefficients to distinguish them from their microphysical counterparts. 
An example of such an effective equation is the mean-field dynamo equation
which, in its simplest form, has two turbulent transport coefficients: the
alpha effect, $\alpha$, and turbulent magnetic diffusivity, $\etat$. 
Once the effective equations and the turbulent transport coefficients are known,
we apply the standard machinery of linear stability theory to the effective equations
to obtain the exponential growth or decay rate of large-scale instabilities in
or even because of the presence of turbulence.

This prescription, applied to real turbulent flows turns out to be not very straightforward
because of several reasons that we list below:
(i) Any spatial averaging procedure will retain some level of fluctuations
\citep{Hoyng}.
This automatically limits the dynamical range over which exponential
growth can be obtained.
The larger the size of the turbulent eddies compared with the size of
the domain, i.e., the smaller the scale separation ratio, the smaller
the dynamical range.
A well-known example is the $\alpha$ effect in mean-field electrodynamics
\citep{Mof78,KR80}, which gives rise to a linear instability of the mean-field
equations.
In direct numerical simulations (DNS), however,
the expected exponential growth can only be seen over a limited dynamical range.
A second, more recent, example is the 
negative effective magnetic pressure instability (NEMPI) \citep{BKKMR11},
where the magnetic pressure develops negative contributions 
caused by the turbulence itself \citep{KRR89,KR94,RK07}.
NEMPI could be detected in DNS only for a scale separation ratio of ten or more. 
(ii) The averaged equation, in addition to the usual diffusive terms,
can have higher order derivatives in both space and time \citep{RB12}. 
Such terms become important for 
a small scale separation ratio that generally reduces the efficiency
of turbulent transport \citep{BRS08,BSV09,MB10}.
So, in general, a simple prescription of replacing the microphysical value
of diffusivity by its turbulent counterpart may not work. 
(iii) There are important conceptual differences between microphysical
and turbulent transport coefficients.  The turbulent ones must reflect the 
anisotropies and inhomogeneities of real flows, and they are hence, in 
general, tensors of rank two or higher. 
Moreover, a major challenge in this formalism is the actual calculation of the
turbulent transport coefficients.  For turbulent flows, there is at present 
no known analytical technique that allows us to calculate them from
first principles.  
A recent breakthrough is the use of the test-field method
\citep{Sch05,Sch07,BRRK08}, which allows us to
numerically calculate the turbulent transport coefficients 
for a large class of flows. 
Armed with the test-field method, we are now in a position to 
quantify how accurately the linear stability theory applied to the
mean-field equations describes the growth of large-scale instabilities in 
a turbulent flow.  This is the principal objective of this paper. 

The magneto-rotational instability (MRI) is a relatively simple
axisymmetric (two-dimensional) linear instability of a rotating shear
flow in the presence of an imposed magnetic field along the rotation axis.
The dispersion relation for MRI is well known. 
Let us now consider the situation in which we have a turbulent flow 
(which may have been generated due to MRI with microphysical parameters)
in a rotating box in the presence of an axial magnetic field and
large-scale shear.  What is the dispersion relation for large-scale
Let us assume that we can use the dispersion relation for MRI and
simply replace the microphysical values of magnetic diffusivity ($\eta$) and
kinematic viscosity ($\nu$) by turbulent values, $\etat$ and $\nut$, respectively.
In that case, the growth rate would be given approximately by
\EQ
\lambda\approx\VA(k) k-(\etat+\eta)k^2,
\label{lambda}
\EN
where $\VA(k)k$ is the growth rate in the non-turbulent, ideal case.
For the MRI with Keplerian shear,
$\VA(k)$ is given in terms of $\tilde\VA=\VA k/\Omega$ with \citep{BH98}
\EQ
\VA(k)^2=\left(\tvA^2+\half\right)
\left\{\left[1+4{(3-\tvA^2)\tvA^2\over(2\tvA^2+1)^2}
\right]^{1/2}\!\!-1\right\},
\EN
where $\tvA=\vA k/\Omega$, $\vA$ is the Alfv\'en speed,
$k$ is the wavenumber, and $\Omega$ is the angular velocity.
The qualitative validity of turbulent diffusion in MRI was previously demonstrated
by \cite{KKV10}, who focussed attention on the Maxwell and Reynolds
stresses in the nonlinear regime, following earlier work by \cite{WA08}
on the combined action of MRI in the presence of forced turbulence.
The effect of forced turbulence on the MRI has been studied previously
in connection with studies of quasi-periodic oscillations driven by the
interaction with rotational and epicyclic frequencies \citep{B05}.
Note also that in \Eq{lambda} we have assumed that
$\nut+\nu = \etat+\eta$ which is essentially equivalent to assuming that the
turbulent magnetic Prandtl number, $\nut/\etat$, is unity because
in most astrophysical flows $\nu\ll\nut$ and $\eta\ll\etat$. 
This assumption is supported by DNS studies \citep{YBR03}. 

There is another important difference between microscopic and turbulent
magnetic diffusion.
For any linear instability the level of the exponentially growing
perturbation depends logarithmically on the strength of the initial field.
However, turbulent diffusion implies the presence of turbulence, so
there is always some non-vanishing projection of the random velocity and
magnetic fields, which will act as a seed such that the growth of the magnetic
field is independent of the initial conditions and depends just on the
value of the forcing wavenumber and the forcing amplitude.
This can become particularly important in connection with the large-scale
dynamo instability, which is an important example of an instability that
operates especially well in a turbulent system.
Again, in that case, turbulence can provide a seed magnetic field
to the large-scale dynamo through the action of the much faster
small-scale dynamo.
This idea was first discussed by \cite{Beck94} in an attempt to
explain the rapid saturation of a large-scale magnetic field in
the galactic dynamo.

\section{Model}

Following earlier work of \cite{WA08} and \cite{KKV10},
we solve the three-dimensional equations of magnetohydrodynamics
(MHD) in a cubic domain of size $L^3$ in the presence of rotation with
angular velocity $\OO=(0,0,\Omega)$, a shear flow $\meanU^S=(0,Sx,0)$
with shear $S=-{3\over2}\Omega$,
and an imposed magnetic field $\BB_0=(0,0,B_0)$.
We adopt shear-periodic boundary conditions in the $x$ direction
\citep{WT88} and periodic boundary conditions in the $y$ and $z$ directions.
We generate turbulence by adding a stochastic force with amplitude $f_0$
and a wavenumber $\kf$.
We have varied $f_0$ to achieve different root-mean-square (rms) velocities of the turbulence.
Different values of the forcing wavenumber $\kf$ will also be considered.

We assume an isothermal gas with sound speed $\cs$, so the pressure
$p=\rho\cs^2$ is linearly related to the density $\rho$.
The hydromagnetic equations are solved in terms of the
magnetic vector potential $\AAA$, the velocity $\UU$, and the
density $\ln\rho$ in the form
\EQ
{\DDD\AAA\over\DDD t}=-SA_y\xxx+\UU\times(\BB+\BB_0)+\eta\nabla^2\AAA,
\EN
\EQA
{\DD\UU\over\DD t}=-SU_x\yyy
+{\JJ\times\BB\over\rho}-c_{\rm s}^2\nab\ln\rho-2\OO\times\UU
\nonumber \\
+\ff
+\nu\left(\nabla^2\UU+\onethird\nab\nab\!\cdot\!\UU
+2\SSSS\nab\ln\rho\right)\!,
\ENA
\EQ
{\DD\ln\rho\over\DD t}=-\nab\cdot\UU,
\EN
where $\DDD/\DDD t=\partial/\partial t+\UU^S\cdot\nab\,$ is
the advective derivative based on the shear flow and
$\DD/\DD t=\DDD/\DDD t+\UU\cdot\nab\,$ is the advective derivative
based on the full flow field that includes
both the shear flow and the deviations from it,
$\BB=\nab\times\AAA$ is the magnetic field expressed in
terms of the magnetic vector potential $\AAA$.
In our units, the vacuum permeability $\mu_0 = 1$.
The current density $\JJ=\nab\times\BB$,
 $\eta$ is the magnetic diffusivity,
$\nu$ is the kinematic viscosity,
and $\ff$ is the turbulent forcing function given by
\EQ
\ff=f_0\mbox{Re}\,\{N\ff_{\kk(t)}\exp[i\kk(t)\cdot\xx+i\phi(t)]\},
\EN
where $f_0$ denotes the non-dimensional forcing amplitude,
\EQ
\ff_{\kk}={\kk\times(\kk\times\eee)-i|\kk|(\kk\times\eee)
\over2\kk^2\sqrt{1-(\kk\cdot\eee)^2/\kk^2}},
\EN
and $\eee$ is an arbitrary unit vector needed to generate
a vector $\kk\times\eee$ that is perpendicular to $\kk$, $\phi(t)$
is a random phase, and $N=f_0 c_{\rm s}(kc_{\rm s}/\delta t)^{1/2}$,
where $f_0$ is a nondimensional factor, $k=|\kk|$, and $\delta t$ is the
length of the time step.
We focus on the case where $|\kk|$ is from a narrow band of wavenumbers
with forcing wavenumber $\kf$.

The smallest wavenumber that fits into the domain is $k_1=2\pi/L$,
and we shall use $k_1$ as our inverse unit length.
Our time unit is given by $\Omega^{-1}$.
Non-dimensional quantities will be expressed by a tilde.
For example the non-dimensional growth rate is $\tilde\lambda=\lambda/\Omega$
and the non-dimensional rms velocity is given by
$\tilde\urms=\urms k_1/\Omega$, and the non-dimensional
Alfv\'en speed is given by $\tvA=\vA k_1/\Omega$,
where $\vA=B_0/\sqrt{\rho_0}$ is the Alfv\'en speed
based on the strength of the imposed magnetic field
and $\rho_0$ is the volume averaged density.
Furthermore, the non-dimensional forcing wavenumber and turbulent
diffusion are given by $\tilde{k}_{\rm f}=k_{\rm f}/k_1$ and
$\tilde\eta_{\rm t}=\eta_{\rm t} k_1^2/\Omega$, respectively.

We quantify our results in terms of fluid and magnetic Reynolds numbers,
as well as the Coriolis number, which are respectively defined as
\EQ
\Rey=\urms/\nu\kf,\quad
\Rm=\urms/\eta\kf,\quad
\Co=2\Omega/\urms\kf.
\label{CoDef}
\EN
In this paper, $\urms$ is the rms velocity before the onset of MRI. 
We characterize our solutions by measuring $\urms$ and a similarly
defined $\brms$, which refers to the departure from the imposed field,
again before the onset of MRI.
We also use the quantity $\Brms$ to characterize the growth of the
total field, given by $\Brms^2 \equiv \brms^2+B_0^2$,
which we use to define the Lundquist number,
\EQ
\Lu=\Brms/\sqrt{\rho}\eta\kf,
\EN
At small magnetic Reynolds numbers, $\Rm\ll1$, we would expect
$\brms/B_0\approx\Rm^{1/2}$ \citep{KR80}, but in most of our runs
we have  $\Rm\gg1$, in which case $\brms/\sqrt{\rho_0}\approx\urms$.
Note that the ratio $\Lu/\Rm$ is then equal to the ratio
of magnetic field to the equipartition value.
We also consider horizontally averaged magnetic field,
$\meanBB(z,t)$, as well as its rms value, which is then still
a function of time.

The DNS are performed with the {\sc Pencil
  Code}\footnote{\url{http://pencil-code.googlecode.com}},
which uses sixth-order explicit
finite differences in space and a third-order accurate
time-stepping method.
We use a numerical resolutions of $128^3$ and $256^3$ mesh points.

In the following, we discuss the dependence of the growth rate on the
anticipated magnetic diffusivity
\EQ
\etatz\equiv\urms/3\kf.
\label{etat}
\EN
This simple formula was previously found to be a good estimate of the
actual value of $\etat$ \citep{SBS08}, but this ignores complications
from a weak dependence on $\kf/k_1$ \citep{BRS08}, as well as the
mean magnetic field \citep{BRRS08},
which would result in magnetic quenching of $\etat$.
To shed more light onto this uncertainty, we also make use of the
quasi-kinematic test-field method of \cite{Sch05,Sch07}
to calculate the actual value
of $\etat$ based on the measured diagonal components of the magnetic
diffusion tensor $\eta_{ij}$, i.e., $\etaT\equiv(\eta_{11}+\eta_{22})/2$.
We recall that the evolution of the horizontally averaged magnetic field
is governed by just four components of $\eta_{ij}$ \citep{BRRK08} and
another four components of what is called the $\alpha_{ij}$ tensor, whose
components turn out to be zero in all cases investigated in this paper.

\section{Results}

\subsection{Turbulence as a seed of MRI}

\begin{figure}[h!]\begin{center}
\includegraphics[width=\columnwidth]{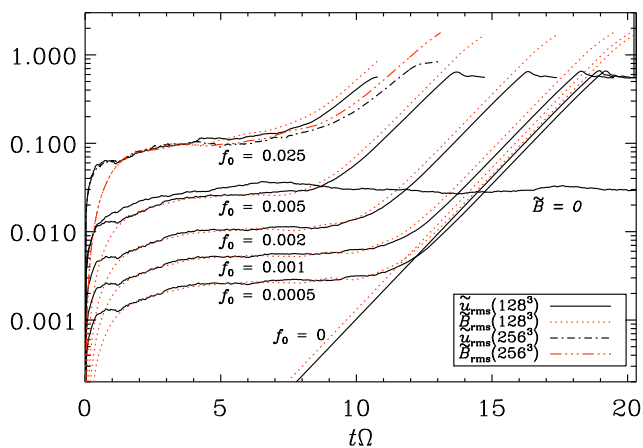}
\end{center}\caption[]{
Time dependence of $\tilde\urms$ and $\tilde\Brms$ for runs with 
$\tilde\kf=2.2$ (Runs~G1-G4 and G7). 
In addition, sample runs with higher resolution ($256^3$),
no magnetic fields, and no forcing are included for comparison. 
}\label{pcomp}\end{figure}

Let us begin by calculating the growth rate of the large-scale
instability from our DNS. The DNS is started with an initial
condition where the velocity is initially zero. 
As a result of the action of the external force, small-scale
velocity grows fast and then saturates. This small-scale velocity
acts as a seed field for the large-scale MRI.
Consequently we see a second growth phase at late times.
This is due to the growth (via MRI) of large-scale velocity and
magnetic field, both of which show exponential growth at
this phase. In \Fig{pcomp} we show this growth for different
values of the amplitude of the external force. 
The growth rate of the large-scale instability can be calculated
from the exponentially growing part of these plots.

At late times, $\tilde\urms$ saturates near unity,
while $\tilde\Brms$ continues to grow; cf.\ \Fig{pcomp}.
Eventually, however, our DNS crash, which is a result
of insufficient resolution.
Increasing the resolution, we have been able to continue the saturated phase
for a somewhat longer time.
The results of a higher resolution run are shown as a long-dashed
line in \Fig{pcomp}, where we used $256^3$ mesh points.
On the other hand, higher resolution is not crucial for determining
the turbulence effects on the MRI, which is why in the following
we only present results obtained at a resolution of $128^3$ mesh points.

\begin{figure}[h!]\begin{center}
\includegraphics[width=\columnwidth]{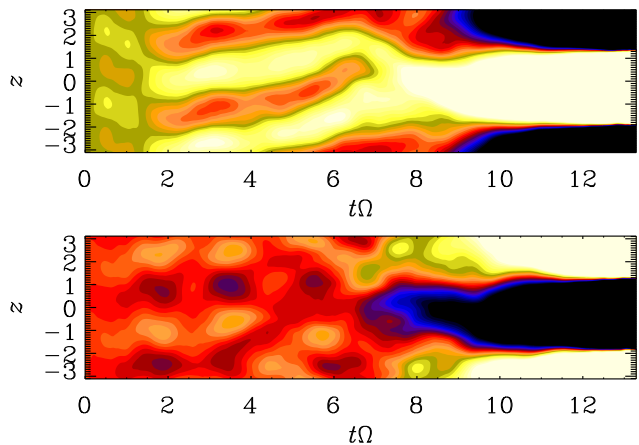}
\end{center}\caption[]{
Time dependence of $\ln\meanB_x$ (top) and $\ln\meanB_y$ (bottom) for
$\tvA = 1.1$ and $\tilde\eta=0.01$ (Run~O7).
Note the different effects of shear on $\meanB_x$ and $\meanB_y$.
}\label{xyaver}\end{figure}

Compared to the non-magnetic case, the magnetic field slightly 
decreases the saturation level of the forced turbulence before the visible growth 
of MRI.
In addition, the presence of a magnetic field causes $\tilde\urms$ to have two 
plateaus: first at the very beginning and second after $\tilde\Brms$ reaches 
the level of $\tilde\urms$. The difference can also be seen in averages presented 
in the \Fig{xyaver}, where polarity of the $\meanB_x$ turns to opposite between 
the first and second plateau.

Once there is exponential growth, the growth rates of
$\urms$ and $\Brms$ are, as expected, the same, but they
are different for different amplitudes $f_0$ of the forcing function,
see also Table~\ref{TSetAll}.
Note also that the runs with the weakest forcing have a slightly
faster growth, because the resulting turbulent viscosity and diffusivity
are smaller, but they also show a later onset of exponential growth.
This in turn is related to a weaker residual projection onto the MRI
eigenfunction, simply because the amplitude of the turbulence is lower.

% fig 1

\begin{figure}[h!]\begin{center}
\includegraphics[width=\columnwidth]{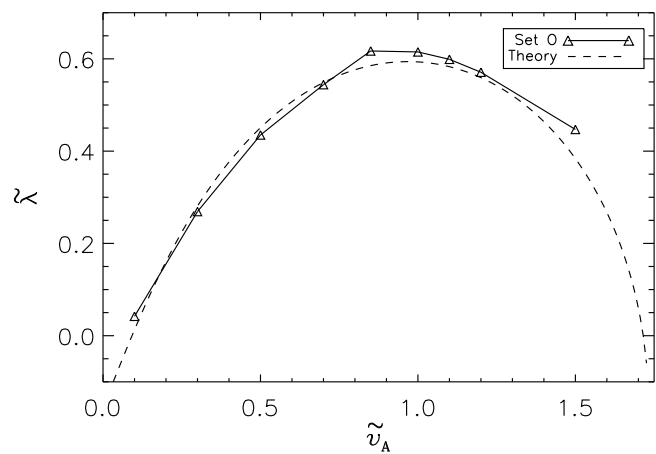}
\end{center}\caption[]{
Dependence of $\tilde\lambda$ on $\tvA$ for the Set~O.
The dashed line represents the dispersion relation of \Eq{lambda} with
$\tilde\etafix = \tilde\eta + \tilde\eta_t = 0.146$.
}\label{presult_vA_lam_Co1}\end{figure}

The growth rate $\lambda$ thus calculated is plotted in \Fig{presult_vA_lam_Co1}
as a function of non-dimensionalized $\vA$.  For comparison, we have
also plotted the growth rate calculated from the dispersion
relation of MRI, \Eq{lambda}, with a fixed coefficient of magnetic diffusivity $\tilde\etafix$,  
where $\tilde\etafix = \tilde\eta + \tilde\etat$. 
We chose a value for $\etafix$ from Run~O3 with $\tvA = 0.50$, 
$\tilde\etafix = 0.01 + 0.136 = 0.146$ (see \Tab{TSetAll}).
Both computed runs and the dispersion relation agree reasonably everywhere 
except with $\tvA > 1.75$, where linear theory predicts no MRI. The
positive growth rates in the DNS results in this regime are likely due
to another instability such as the incoherent $\alpha$--shear dynamo
\citep{VB97,MB12} and/or the turbulent shear
dynamo \citep{YHSea08a,YHSea08b,Heine11}.

% fig 2

In the beginning, the components of the horizontally averaged magnetic
field $\meanBB$
are still randomly fluctuating, but at later times, when nonlinear effects
begin to play a role, a clear pattern with
wavenumber $k=k_1$ develops; see \Fig{xyaver}.
This is expected in this particular run (O7) where the fastest growing
mode has a wavenumber close to $k_1$. However, we see the same
behavior also in other runs in Set~O where the theoretically predicted
$k_{\rm max}$ varies by more than an order of magnitude, see
\Tab{TSetAll}.
By contrast, according to linear theory, the eigenfunction always
settles onto the fastest growing one, which would have a wavenumber larger
than $k_1$.

% fig 3

\subsection{Different ways of varying $\etat$}

To explore the dependence of the solutions on the anticipated turbulent
magnetic diffusivity $\etat$, we consider three sets of runs.
In two of them (Sets~A and B), we vary $\kf$, and in one (Set~C) we vary the value of $\Omega$,
thus changing $\Co$ which was defined in \Eq{CoDef}.
Given the definition of $\etatz$ in \Eq{etat}, we have
\begin{equation}
\tilde{\eta}_{\rm t0}={1\over3}\tilde\urms/\tilde\kf
={2\over3}\left({\kf\over k_1}\right)^{-2}\,\Co^{-1}.
\label{tildeetat}
\end{equation}
This shows that increasing either $\Co$ or $\kf$ or both
leads to a decrease of $\tilde\etatz$.
We recall that $\tilde\urms$ is the value before the onset of MRI
and has been estimated by measuring the height of the plateau
seen in \Fig{pcomp}.
We should point out that for small values of $\tilde\kf$
the length of the plateau becomes rather short, which leads
therefore to a significant source of error.
The parameters for the three sets of runs are summarized in
\Tab{TSetAll}.

% fig 4
\begin{figure}[h!]\begin{center}
\includegraphics[width=\columnwidth]{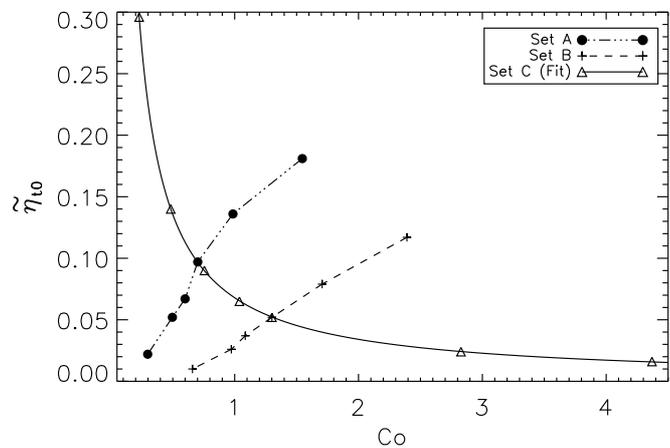}
\end{center}\caption[]{
$\tilde\etatz$ versus Coriolis number for $\tvA=1$,
at a resolution of $128^3$ mesh points.
%The solid line is a fit of Eq. \ref{tildeetat} into the results of Set~C. 
%AB: use macro (need brackets)
The solid line is a fit of \Eq{tildeetat} into the results of Set~C. 
}\label{presult_co_tetat}\end{figure}

\begin{figure}[h!]\begin{center}
\includegraphics[width=\columnwidth]{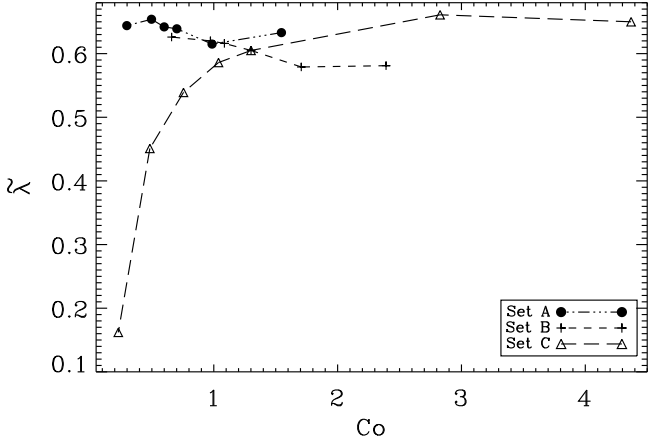}
\end{center}\caption[]{
Growth rate versus Coriolis number for $\tvA=1$
at a resolution of $128^3$ mesh points.
}\label{presult_coriolis}\end{figure}

In \Fig{presult_co_tetat} we plot these three sets of runs in a
$\Co$--$\tilde\etatz$ diagram.
Looking at \Eq{tildeetat}, and since $\tilde\kf=\kf/k_1=2.3$ is fixed,
it is clear that the runs of Set~C all fall on a line proportional
to $\Co^{-1}$.
For the other two sets, $\tilde\kf$ varies.
Small values of $\tilde\kf$ correspond to large values of both
$\Co$ and $\tilde\etatz$, and vice versa, which is the reason why
the other two branches for Sets~A and B show an increase of
$\tilde\etatz$ for increasing values of $\Co$.
Correspondingly, $\tilde\lambda$ decreases with increasing $\Co$
for Sets~A and B, while for Set~C, $\tilde\lambda$ increases with
increasing $\Co$; see \Fig{presult_coriolis}.

% fig 5

\begin{figure}[h!]\begin{center}
\includegraphics[width=\columnwidth]{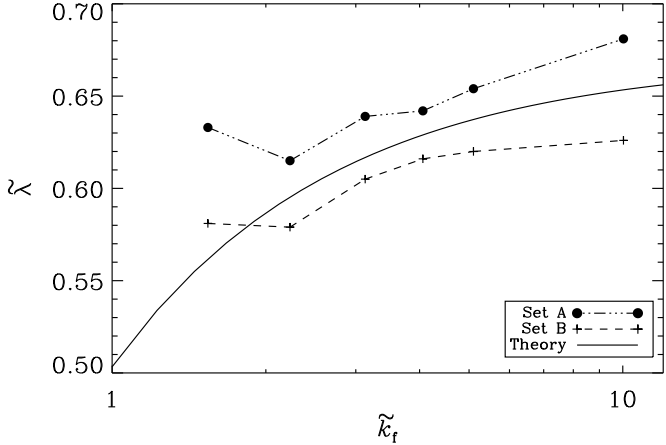}
\end{center}\caption[]{
Growth rate versus the scale separation $\tilde\kf$ for
$\tvA=1$, at a resolution of $128^3$ mesh points.
The solid line shows a fit to the theoretical dependency given by
\Eq{lam_vs_kf}.
}\label{presult_kf}\end{figure}

For Sets~A and B we show the dependence of the growth rate on $\tilde\kf$
in \Fig{presult_kf}.
For both sets, $\tilde\lambda$ increases with increasing $\tilde\kf$.
This increase is related to the fact for increasing values of $\tilde\kf$,
$\tilde\etatz$ decreases, and thus $\tilde\lambda$ shows a mild increase.
Indeed, we should expect that $\tilde\lambda$ varies with $\tilde\kf$ like
\EQ
\tilde\lambda=\tilde\lambda_0-\tilde\urms/3\tilde\kf,
\label{lam_vs_kf}
\EN
where in the present case the best agreement with the DNS is obtained
when $\tilde\lambda_0=0.67$ is chosen.
This theoretically expected dependency is overplotted in \Fig{presult_kf}.

% fig 6

% fig 7
\begin{figure}[h!]\begin{center}
\includegraphics[width=\columnwidth]{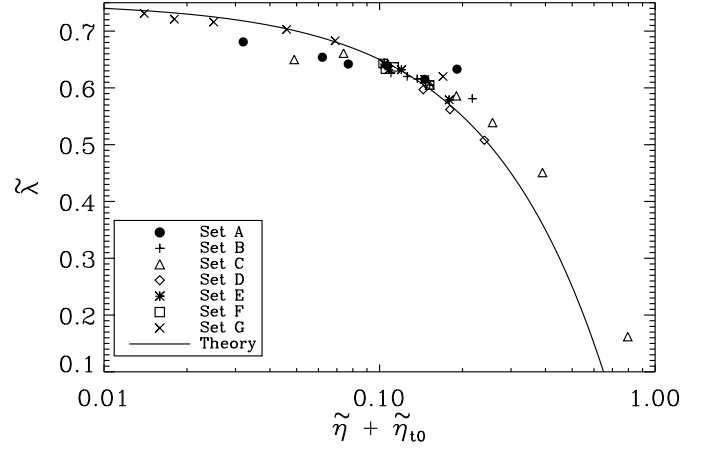}
\end{center}\caption[]{
Growth rate versus the $\tilde{\eta}_{\rm t0}$ for Sets~A--G,
at a resolution of $128^3$ mesh points. 
}\label{presult_tetat}\end{figure}

Let us now turn to relation \eq{lambda}, which predicts a parabolic
decline for increasing values of $(\etat+\eta)k^2$.
This relation is surprisingly well obeyed; see \Fig{presult_tetat},
where we plot $\tilde\lambda$ as a function of $(\etat+\eta)k^2$ for
models of all three sets, together with those of Sets~D--G
listed in \Tab{TSetAll}.

\subsection{Comparison with test-field results}

Our results presented so far have demonstrated that in the present problem, 
the growth of large-scale perturbations is determined by the same equations
that describe the growth of MRI but with values of magnetic diffusivity (and viscosity)
that are not their microphysical values but turbulent values.  
Hence, by turning the problem on its head, we have here a new method of calculating
the turbulent magnetic diffusivity by measuring the growth rate of the
large-scale instability. 
Such a method would proceed in the following manner. 
First we would study the growth of the large-scale instability and produce a plot
similar to \Fig{pcomp} from which we can calculate the growth rate $\lambda$. 
Once we know $\lambda$ we can read off $\etat$ by using  \Fig{presult_tetat}. 
Let us call the turbulent diffusivity, measured in this fashion, $\etatMRI$. 
At present, we are already familiar with the well-established test-field
method to calculate the turbulent magnetic diffusivity.
It then behooves us to compare these two methods, for cases where they
both can be applied.

% fig 8
\begin{figure}[t!]\begin{center}
\includegraphics[width=\columnwidth]{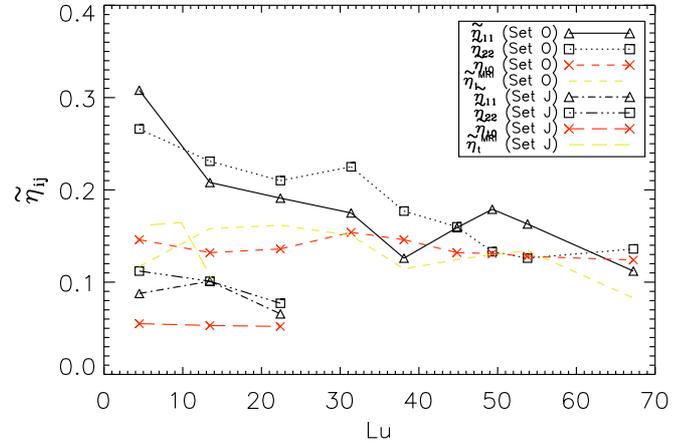}
\end{center}\caption[]{
Dependence of $\tilde\eta_{11}$ ($\bigtriangleup$) and $\tilde\eta_{22}$ ($\square$)
as a function of Lu compared to $\tilde\etatz$ and $\etatMRI$. 
}\label{presult_vA_etatensor}\end{figure}

% fig 9
\begin{figure}[t!]\begin{center}
\includegraphics[width=\columnwidth]{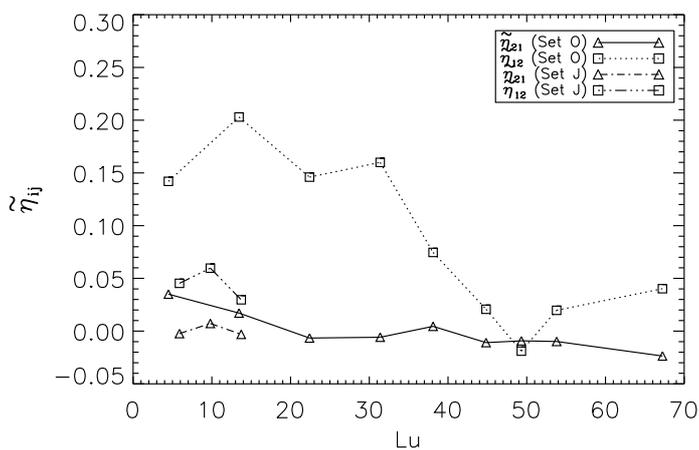}
\end{center}\caption[]{
Dependence of $\tilde\eta_{12}$ ($\square$) and $\tilde\eta_{21}$ ($\bigtriangleup$)
as a function of Lu.
}\label{presult_vA_etatensor_delta}\end{figure}

To apply the test-field method to the present problem, we define averaged quantities by
averaging over the horizontal $xy$ plane and choose $z$-dependent test fields
which are sines and cosines.  
In principle the turbulent magnetic diffusivity thus calculated is a second-rank tensor, 
$\eta_{ij}$.  We plot the diagonal and off-diagonal components of this tensor
in \Figs{presult_vA_etatensor}{presult_vA_etatensor_delta}, respectively.
The off-diagonal elements are close to zero and the diagonal elements
are equal to each other and also equal to $\etatz$.
In \Tab{TSetAll} we list $\tilde\etaT=(\tilde\eta_{11}+\tilde\eta_{22})/2$.
Regarding the off-diagonal elements,
if any departure from zero is significant, it would be for small values
of $\tvA$, i.e., in the kinematic regime where the effects of magnetic
quenching are weak.
Earlier DNS of \cite{B05} of MRI in the presence of forcing
also showed that $\eta_{12}$ was larger than $\eta_{21}$ by magnitude,
but their signs were opposite to those found here.
The reason for this difference is however unclear.

\subsection{Is there $\etat$ quenching?}

The two methods we have described and compared in the previous subsection
now allow us to quantify how turbulent
diffusivity is quenched in the presence of the background magnetic field.
Quenching of turbulent magnetic diffusivity has been computed analytically
\citep{KRP94} and numerically \citep{YBR03}, and it has been used in
dynamo models \citep{Tob96,Gue09}.
Here, we address this issue by considering the turbulent magnetic diffusivity
$\etatMRI$ and $\etatz$ as a function of $\Lu$, as done in
\Fig{presult_vA_etatensor}.
In none of the cases do we observe any $\etat$ quenching.

For Set~G we see that $\etatz$ shows an increase with magnetic field
strength (see \Tab{TSetAll}),
which might suggest the possibility of ``anti-quenching''.
However, in Set~G, the value of $\Rm$ is also increasing, so the increase
in $\etaT$ is really just a consequence of too small values of $\Rm$ in
the runs with weak magnetic field.
This is confirmed by considering the runs in Set~O, where $\Rm$ is
approximately constant and $\etaT$ is then found to be approximately
independent of the imposed field strength.
It should however be pointed out that the possibility of
anti-quenching of $\etat$
(as well as anti-quenching of the $\alpha$ effect in dynamo theory)
has been invoked in the past to explain the observed increase of the
ratio of dynamo frequency to rotational frequency for more active stars
\citep{BST98}.
Anti-quenching of $\etat$ and $\alpha$  was also found
for flows driven by the magnetic buoyancy instability \citep{CMRB11}.
On the other hand, regular quenching has been found both in the absence of
shear \citep{BRRS08} as well as in the presence of shear \citep{KB09}.
It should therefore be checked whether earlier findings of anti-quenching
may also have been affected by too small magnetic Reynolds numbers.

\section{Conclusion}

Our work has demonstrated several unexpected aspects of turbulent
mixing on the operation of the MRI.
Firstly, the effect of turbulent magnetic diffusivity seems to be in
all aspects equivalent to that of microphysical magnetic diffusivity.
This is true even when scale separation is poor, e.g., for
$\kf/k_1=1.5$ or 2.2.
This is rather surprising, because in such an extreme case the memory
effect was previously found to be important \citep{BKM04}, which means
that higher time derivatives in the mean-field parameterization need
to be included \citep{HB09}.
Secondly, the simple estimate given by \Eq{etat} is remarkably accurate.
As a consequence, \Eq{lambda} provides a quantitatively useful estimate
for the effects of turbulence on the growth rate of the MRI.
Our simple estimates also agree with results obtained from the test-field
method.
In principle, there could be other non-diffusive effects resulting
from the so-called $\OO\times\JJ$ effect \citep{Rad69} or the shear--current
effect \cite{RK03,RK04}, but our present results show that this does not seem
to be the case.

It should also be pointed out that no new terms seem to appear in the
momentum equation other than the turbulent viscous force.
Of course, this could change if we were to allow for extra effects
such as strong density stratification, which could lead to the development
of the negative effective magnetic pressure instability
\citep[see][and references therein]{BKKMR11}.
Furthermore, if there is cross-helicity, there can be new
terms in the momentum equation
that are linear in the mean magnetic field \citep{RB10}.
Also kinetic and magnetic helicity could affect our results, although
there have not yet been any indications for this from purely hydrodynamic
shear flow turbulence \citep{MB10}.
Neither the the negative effective magnetic pressure instability nor the
$\alpha$ effect dynamo instability are possible in the simple example
studied here, because stratification is absent.
However, as alluded to in the introduction, they both are examples that
have contributed to the motivation of the work presented here.

%Acknowledgements 

\begin{acknowledgements}
  The authors thank Nordita for hospitality during their visits. 
  Financial support from a Jenny and Antti Wihuri Foundation grant
  (MV), the Academy of Finland grants No.\ 136189, 140970 (PJK) and
  218159, 141017 (MJM), as well as the Swedish Research Council grants
  621-2011-5076 and 2012-5797, and the European Research Council under
  the AstroDyn Research Project 227952 are acknowledged.
  We acknowledge CSC -- IT Center for Science Ltd., who are
  administered by the Finnish Ministry of Education, for the
  allocation of computational resources.
  This research has made use of NASA's Astrophysics Data System.

\end{acknowledgements}

%r e f

\begin{appendix}

\section{Online material}

\begin{table*}[p]
\scriptsize 
\caption{Results for all datasets. 
Each dataset have been labeled with a specific letter. Some sets share 
single DNS runs with each other, which have been marked on the table. 
}\centerline{\begin{tabular}{lrcccccccccccccc}
Run & $\tilde{f}_0$ & $\tilde\kf$ & $\kmax$ & $\tvA$ &   $\Rm$  &    $\Co$ &  $\Lu$ & $\Omega$ & $\tilde\eta$   & 
$\tilde\etatz$ & $\tilde\etatMRI$ & $\tilde\lambda$ & $\tilde\urms$ & $\tilde\brms$ & $\tilde\etaT$ \\
\hline
N1 & 0.0000 &    -   &     1.0 &    1.00 &    -     &  -  &   - &     0.10 &  0.010 &  - &  -  &  0.739 &  - &  - & - \\% 128_vA010_eta0001_Om01_noforce
N2 & 0.0000 &    -   &     0.9 &    1.10 &    -     &  -  &   - &     0.10 &  0.010 &  - &  -  &  0.729 &  - &  - & - \\% T128_vA011_eta0001_Om01_noforce_xyaver
\hline
O1 & 0.0200 &    2.2 &   10.0 &    0.10 &   43.79 &     0.92 &     4.48 &     0.10 &  0.010 &  0.146 &  0.117 &  0.042 &  0.977 &  0.843 & 0.287 \\% T128_vA001_eta0001_Om01_k2_f_0020_xyaver
O2 & 0.0200 &    2.2 &    3.3 &    0.30 &   39.59 &     1.02 &    13.45 &     0.10 &  0.010 &  0.132 &  0.158 &  0.269 &  0.883 &  0.797 & 0.219 \\% T128_vA003_eta0001_Om01_k2_f_0020_xyaver
O3 & 0.0200 &    2.2 &    2.0 &    0.50 &   40.70 &     0.99 &    22.41 &     0.10 &  0.010 &  0.136 &  0.162 &  0.435 &  0.908 &  0.973 & 0.200 \\% T128_vA005_eta0001_Om01_k2_f_0020_xyaver
O4 & 0.0200 &    2.2 &    1.4 &    0.70 &   46.35 &     0.87 &    31.38 &     0.10 &  0.010 &  0.154 &  0.150 &  0.544 &  1.034 &  1.161 & 0.200 \\% T128_vA007_eta0001_Om01_k2_f_0020_xyaver
O5 & 0.0200 &    2.2 &    1.2 &    0.85 &   43.78 &     0.92 &    38.10 &     0.10 &  0.010 &  0.146 &  0.114 &  0.617 &  0.977 &  1.038 & 0.151 \\% T128_vA0085_eta0001_Om01_k2_f_0020_xyaver
O6 & 0.0200 &    2.2 &    1.0 &    1.00 &   39.72 &     1.01 &    44.83 &     0.10 &  0.010 &  0.132 &  0.125 &  0.615 &  0.886 &  0.881 & 0.160 \\% T128_vA010_eta0001_Om01_k2_f_0020
O7 & 0.0200 &    2.2 &    0.9 &    1.10 &   39.33 &     1.02 &    49.31 &     0.10 &  0.010 &  0.131 &  0.129 &  0.599 &  0.877 &  0.850 & 0.156 \\% T128_vA011_eta0001_Om01_k2_f_0020_xyaver
O8 & 0.0200 &    2.2 &    0.8 &    1.20 &   38.30 &     1.05 &    53.79 &     0.10 &  0.010 &  0.128 &  0.135 &  0.571 &  0.854 &  0.804 & 0.144 \\% T128_vA012_eta0001_Om01_k2_f_0020_xyaver
O9 & 0.0200 &    2.2 &    0.7 &    1.50 &   37.21 &     1.08 &    67.24 &     0.10 &  0.010 &  0.124 &  0.083 &  0.447 &  0.830 &  0.737 & 0.124 \\% T128_vA015_eta0001_Om01_k2_f_0020_xyaver
O10 & 0.0200 &    2.2 &    0.6 &    1.75 &   39.55 &     1.02 &    78.45 &     0.10 &  0.010 &  0.132 &   --- &  0.032 &  0.882 &  0.771 & 0.155 \\% T128_vA0175_eta0001_Om01_k2_f_0020_xyaver
O11 & 0.0200 &    2.2 &    0.5 &    2.00 &   40.21 &     1.00 &    89.65 &     0.10 &  0.010 &  0.134 &   --- &  0.002 &  0.897 &  0.759 & 0.142 \\% T128_vA020_eta0001_Om01_k2_f_0020_xyaver
O12 & 0.0200 &    2.2 &    0.4 &    2.50 &   40.19 &     1.00 &   112.07 &     0.10 &  0.010 &  0.134 &   --- &  0.001 &  0.897 &  0.752 & 0.168 \\% T128_vA025_eta0001_Om01_k2_f_0020_xyaver
\hline 
A1 & 0.0200 &    1.5 &    1.0 &    1.00 &   54.45 &     1.55 &    64.88 &     0.10 &  0.010 &  0.181 &  0.107 &  0.633 &  0.839 &  0.785 & - \\% 128_vA010_eta0003_Om01_k1
A2 & 0.0200 &    2.2 &    1.0 &    1.00 &   40.76 &     0.99 &    44.83 &     0.10 &  0.010 &  0.136 &  0.125 &  0.615 &  0.909 &  0.918 & - \\% 128_vA010_eta0003_Om01_k2
A3 & 0.0200 &    3.1 &    1.0 &    1.00 &   29.02 &     0.70 &    31.91 &     0.10 &  0.010 &  0.097 &  0.100 &  0.639 &  0.909 &  0.927 & - \\% 128_vA010_eta0003_Om01_k3
A4 & 0.0200 &    4.1 &    1.0 &    1.00 &   20.24 &     0.60 &    24.63 &     0.10 &  0.010 &  0.067 &  0.098 &  0.642 &  0.822 &  0.767 & - \\% 128_vA010_eta0003_Om01_k4
A5 & 0.0200 &    5.1 &    1.0 &    1.00 &   15.49 &     0.50 &    19.62 &     0.10 &  0.010 &  0.052 &  0.085 &  0.654 &  0.789 &  0.750 & - \\% 128_vA010_eta0003_Om01_k5
A6 & 0.0200 &   10.0 &    1.0 &    1.00 &    6.68 &     0.30 &     9.97 &     0.10 &  0.010 &  0.022 &  0.058 &  0.681 &  0.670 &  0.559 & - \\% 128_vA010_eta0003_Om01_k10
\hline
B1 & 0.0200 &    1.5 &    1.0 &    1.00 &    3.52 &     2.39 &     6.49 &     0.10 &  0.100 &  0.117 &  0.069 &  0.581 &  0.543 &  0.453 & - \\% 128_vA010_eta003_Om01_k1
B2 & 0.0200 &    2.2 &    1.0 &    1.00 &    2.36 &     1.71 &     4.48 &     0.10 &  0.100 &  0.079 &  0.070 &  0.579 &  0.525 &  0.425 & - \\% 128_vA010_eta003_Om01_k2
B3 & 0.0200 &    3.1 &    1.0 &    1.00 &    1.57 &     1.30 &     3.19 &     0.10 &  0.100 &  0.052 &  0.044 &  0.605 &  0.491 &  0.342 & - \\% 128_vA010_eta003_Om01_k3
B4 & 0.0200 &    4.1 &    1.0 &    1.00 &    1.12 &     1.09 &     2.46 &     0.10 &  0.100 &  0.037 &  0.034 &  0.616 &  0.454 &  0.284 & - \\% 128_vA010_eta003_Om01_k4
B5 & 0.0200 &    5.1 &    1.0 &    1.00 &    0.79 &     0.97 &     1.96 &     0.10 &  0.100 &  0.026 &  0.029 &  0.620 &  0.403 &  0.225 & - \\% 128_vA010_eta003_Om01_k5
B6 & 0.0200 &   10.0 &    1.0 &    1.00 &    0.30 &     0.66 &     1.00 &     0.10 &  0.100 &  0.010 &  0.023 &  0.626 &  0.302 &  0.107 & - \\% 128_vA010_eta003_Om01_k10
\hline 
C1 & 0.0200 &    3.1 &    1.0 &    1.00 &    1.40 &     4.37 &     9.57 &     0.30 &  0.033 &  0.016 &  0.066 &  0.650 &  0.146 &  0.122 & - \\% 128_vA030_eta003_Om03_k3
C2 & 0.0200 &    3.1 &    1.0 &    1.00 &    1.44 &     2.83 &     6.38 &     0.20 &  0.050 &  0.024 &  0.038 &  0.661 &  0.226 &  0.192 & - \\% 128_vA020_eta003_Om02_k3
C3 \tablefootmark{a} & 0.0200 &    3.1 &    1.0 &    1.00 &    1.57 &     1.30 &     3.19 &     0.10 &  0.100 &  0.052 &  0.044 &  0.605 &  0.491 &  0.342 & - \\% 128_vA010_eta003_Om01_k3
C4 & 0.0200 &    3.1 &    1.0 &    1.00 &    1.57 &     1.04 &     2.55 &     0.08 &  0.125 &  0.065 &  0.038 &  0.586 &  0.615 &  0.392 & - \\% 128_vA008_eta003_Om008_k3
C5 & 0.0200 &    3.1 &    1.0 &    1.00 &    1.62 &     0.76 &     1.91 &     0.06 &  0.167 &  0.090 &  0.044 &  0.539 &  0.845 &  0.470 & - \\% 128_vA006_eta003_Om006_k3
C6 & 0.0200 &    3.1 &    1.0 &    1.00 &    1.68 &     0.48 &     1.28 &     0.04 &  0.250 &  0.140 &  0.049 &  0.451 &  1.318 &  0.583 & - \\% 128_vA004_eta003_Om004_k3
C7 & 0.0200 &    3.1 &    1.0 &    1.00 &    1.78 &     0.23 &     0.64 &     0.02 &  0.500 &  0.296 &  0.087 &  0.162 &  2.787 &  1.131 & - \\% 128_vA002_eta003_Om002_k3
\hline 
D1 & 0.0200 &    3.1 &    1.0 &    1.00 &    0.09 &     9.41 &     1.37 &     0.30 &  0.233 &  0.007 &  0.008 &  0.508 &  0.068 &  0.030 & - \\% 128_vA030_eta007_Om03_k3
D2 & 0.0200 &    3.1 &    1.0 &    1.00 &    0.09 &    12.71 &     1.82 &     0.40 &  0.175 &  0.005 &  0.012 &  0.562 &  0.050 &  0.027 & - \\% 128_vA040_eta007_Om04_k3
D3 & 0.0200 &    3.1 &    1.0 &    1.00 &    0.09 &    15.82 &     2.28 &     0.50 &  0.140 &  0.004 &  0.013 &  0.597 &  0.040 &  0.023 & - \\% 128_vA050_eta007_Om05_k3
\hline 
E1 & 0.0010 &    2.2 &    1.0 &    1.00 &    0.12 &    33.43 &     4.48 &     0.10 &  0.100 &  0.004 &  0.007 &  0.642 &  0.027 &  0.022 & - \\% 128_vA010_eta001_Om01_k2_f_0001
E2 & 0.0020 &    2.2 &    1.0 &    1.00 &    0.24 &    16.69 &     4.48 &     0.10 &  0.100 &  0.008 &  0.018 &  0.632 &  0.054 &  0.044 & - \\% 128_vA010_eta001_Om01_k2_f_0002
E3 & 0.0050 &    2.2 &    1.0 &    1.00 &    0.59 &     6.76 &     4.48 &     0.10 &  0.100 &  0.020 &  0.018 &  0.632 &  0.133 &  0.109 & - \\% 128_vA010_eta001_Om01_k2_f_0005
E4 \tablefootmark{b} & 0.0200 &    2.2 &    1.0 &    1.00 &    2.36 &     1.71 &     4.48 &     0.10 &  0.100 &  0.079 &  0.070 &  0.579 &  0.525 &  0.425 & - \\% 128_vA010_eta003_Om01_k2
\hline 
F1 & 0.0010 &    3.1 &    1.0 &    1.00 &    0.08 &    25.70 &     3.19 &     0.10 &  0.100 &  0.003 &  0.006 &  0.643 &  0.025 &  0.017 & - \\% 128_vA010_eta001_Om01_k3_f_0001
F2 & 0.0020 &    3.1 &    1.0 &    1.00 &    0.16 &    12.78 &     3.19 &     0.10 &  0.100 &  0.005 &  0.017 &  0.633 &  0.050 &  0.033 & - \\% 128_vA010_eta001_Om01_k3_f_0002
F3 & 0.0050 &    3.1 &    1.0 &    1.00 &    0.39 &     5.17 &     3.19 &     0.10 &  0.100 &  0.013 &  0.012 &  0.637 &  0.124 &  0.084 & - \\% 128_vA010_eta001_Om01_k3_f_0005
F4 \tablefootmark{c} & 0.0200 &    3.1 &    1.0 &    1.00 &    1.57 &     1.30 &     3.19 &     0.10 &  0.100 &  0.052 &  0.044 &  0.605 &  0.491 &  0.342 & - \\% 128_vA010_eta003_Om01_k3
\hline 
G1 & 0.0005 &    2.2 &    1.0 &    1.00 &    1.16 &    34.54 &    44.83 &     0.10 &  0.010 &  0.004 &  0.008 &  0.731 &  0.026 &  0.025 & - \\% 128_vA010_eta0001_Om01_k2_f_00005
G2 & 0.0010 &    2.2 &    1.0 &    1.00 &    2.33 &    17.27 &    44.83 &     0.10 &  0.010 &  0.008 &  0.019 &  0.721 &  0.052 &  0.051 & - \\% 128_vA010_eta0001_Om01_k2_f_0001
G3 & 0.0020 &    2.2 &    1.0 &    1.00 &    4.56 &     8.82 &    44.83 &     0.10 &  0.010 &  0.015 &  0.024 &  0.716 &  0.102 &  0.099 & - \\% 128_vA010_eta0001_Om01_k2_f_0002
G4 & 0.0050 &    2.2 &    1.0 &    1.00 &   10.88 &     3.69 &    44.83 &     0.10 &  0.010 &  0.036 &  0.036 &  0.703 &  0.243 &  0.235 & - \\% 128_vA010_eta0001_Om01_k2_f_0005
G5 & 0.0080 &    2.2 &    1.0 &    1.00 &   17.69 &     2.27 &    44.83 &     0.10 &  0.010 &  0.059 &  0.057 &  0.683 &  0.395 &  0.390 & - \\% 128_vA010_eta0001_Om01_k2_f_0008
G6 \tablefootmark{d} & 0.0200 &    2.2 &    1.0 &    1.00 &   40.76 &     0.99 &    44.83 &     0.10 &  0.010 &  0.136 &  0.125 &  0.615 &  0.909 &  0.918 & - \\% 128_vA010_eta0003_Om01_k2
G7 & 0.0250 &    2.2 &    1.0 &    1.00 &   47.92 &     0.84 &    44.83 &     0.10 &  0.010 &  0.160 &  0.120 &  0.620 &  1.069 &  1.086 & - \\% 128_vA010_eta0001_Om01_k2_f_0025
\hline 
J1 & 0.0200 &    5.1 &    3.3 &    0.30 &   16.61 &     0.46 &     5.89 &     0.10 &  0.010 &  0.055 &  0.162 &  0.265 &  0.847 &  0.707 & 0.100 \\% T128_vA003_eta0001_Om01_k5_f_0020_xyaver
J2 & 0.0200 &    5.1 &    2.0 &    0.50 &   15.76 &     0.49 &     9.81 &     0.10 &  0.010 &  0.053 &  0.165 &  0.432 &  0.803 &  0.718 & 0.101 \\% T128_vA005_eta0001_Om01_k5_f_0020_xyaver
J3 & 0.0200 &    5.1 &    1.4 &    0.70 &   15.53 &     0.50 &    13.73 &     0.10 &  0.010 &  0.052 &  0.102 &  0.592 &  0.791 &  0.743 & 0.071 \\% T128_vA007_eta0001_Om01_k5_f_0020_xyaver
\label{TSetAll}\end{tabular}}
\tablefoottext{a}{Same as B3.}
\tablefoottext{b}{Same as B2.}
\tablefoottext{c}{Same as B3 and C3.}
\tablefoottext{d}{Same as A2.}
\end{table*}

\end{appendix}

%\vfill\bigskip\noindent\tiny\begin{verbatim}
%$Header: /var/cvs/brandenb/tex/mvaisala/mri/paper.tex,v 1.157 2013-10-11 13:23:11 mvaisala Exp $

%\end{verbatim}

\end{document}